\documentclass[reprint,showpacs]{revtex4-1}

\usepackage{amssymb}

\usepackage{amsfonts}

\usepackage{latexsym}

\usepackage{dcolumn}

\usepackage{amsmath}

\usepackage{graphicx}

\usepackage{psfrag,pstricks}

\usepackage[cm]{fullpage}
\usepackage{multirow}

\begin{document}

\title{Phonon Josephson Junction with Nanomechanical Resonators}

\author{Shabir Barzanjeh$^{1,2}$}
\author{David Vitali$^{3}$}
\affiliation{$^{1}$Institute of Science and Technology Austria, 3400 Klosterneuburg, Austria}
\affiliation{$^{2}$JARA-Institute for Quantum Information, RWTH Aachen University, 52056 Aachen, Germany}

\affiliation{$^{3}$Physics Division, School of Science and Technology, University of
Camerino, I-62032 Camerino, Italy, and INFN, Sezione di Perugia, Perugia, Italy}
\date{\today}

\begin{abstract}
We study coherent phonon oscillations and tunneling between two coupled nonlinear nanomechanical resonators. We show that the coupling between two nanomechanical resonators creates an effective phonon Josephson junction which exhibits two different dynamical behaviors: Josephson oscillation~(phonon-Rabi oscillation) and macroscopic self-trapping~(phonon blockade). Self-trapping originates from mechanical nonlinearities, meaning that when the nonlinearity exceeds its critical value, the energy exchange between the two resonators is suppressed, and phonon-Josephson oscillations between them are completely blocked. An effective classical Hamiltonian for the phonon Josephson junction is derived and its mean-field dynamics is studied in phase space. Finally, we study the phonon-phonon coherence quantified by the mean fringe visibility, and show that the interaction between the two resonators may lead to the loss of coherence in the phononic junction.
\end{abstract}
\maketitle
\section{introduction}
Nano-electromechanical and optomechanical resonators~\cite{Blencowe,Favero,amo,Eom,Meystre,rmp} are widely used structures, which can be applied for the sensitive detection of physical quantities such as spin~\cite{spin1,spin2}, atomic/molecular mass~\cite{mol1,mol2,mol3}, biological samples~\cite{Shekhawat}, thermal fluctuation~\cite{Badzey,Paul,Tamayo} and also for testing quantum mechanics at the macroscopic level~\cite{O'Connell,Teufel,Poot,Pikovski,Bassi,Bawaj,Liberati} or frequency conversion~\cite{Barzanjeh2011,Barzanjeh1,Andrews, Barzanjeh2}. Nanomechanical resonators~(NMRs) with resonance frequencies in the GHz regime can be now fabricated~\cite{O'Connell,Huang,Peng,Painter1} and this makes them suitable candidates for the study of the quantum behavior at the mesoscopic scale~\cite{O'Connell,Painter1,Painter2}. These GHz NMRs are characterized by reduced dimensions and therefore by very low masses, and at the same time, in this regime the nonlinear behavior of the mechanical systems becomes more relevant, consequently offering interesting theoretical~\cite{Lifshitz,Cross,Katz, Barzanjeh3,Rips,Rips1} and experimental challenges~\cite{Yu,Aldridge,Kozinsky}. These high-frequency resonators operating at nonlinear regime open up new possibilities for the realization of novel devices and applications of NMR and nanoelectromechanical systems~\cite{Lupascu,Woolley}.

In this paper, we explore the link between the enforcement of nonlinearity in two coupled nanomechanical resonators and the emergence of Josephson junction like interaction in the dynamical behavior of mechanical systems. In particular, we explore the coherent phonon oscillations and tunneling between two coupled nonlinear mechanical resonators and show that the mode coupling between two nanomechanical resonators introduces a novel phenomenon, an effective \textit{phonon Josephson junction}, that is a phononic analog of the superconducting Josephson junction~(SJJ) and of the bosonic Josephson junction, which has been proposed theoretically~\cite{Smerzi,Raghavan,Gati,Boukobza,Naether,Szirmai2} and realized experimentally~\cite{Wang,Albiez,Levy,Estve,Zibold,LeBlanc} with ultra-cold atoms trapped within double-well potentials or optical lattices. We also mention that a photonic analog of the Josephson effect in two weakly linked microcavities has been investigated in~\cite{Ji,Larson,Teng}.
\begin{figure}[ht]
\centering
\includegraphics[width=2.5in]{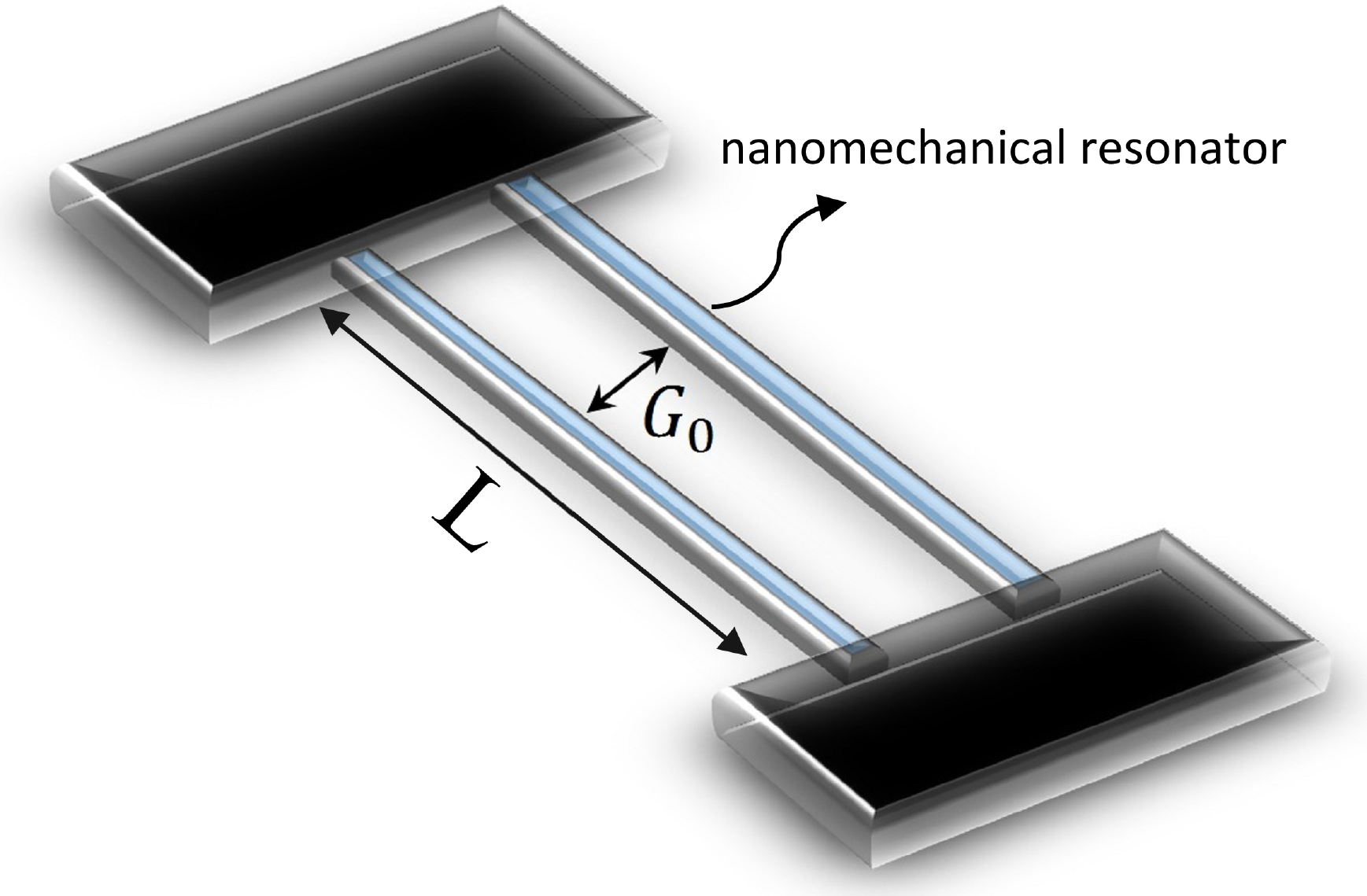}
\caption{Schematic of two coupled nonlinear nanomechanical resonators, here in the form of doubly
clamped beams of length $L$, vibrating in their fundamental flexural mode $ \omega_{0i } $. The mechanical-mechanical coupling rate is $ G_0 $. }
\label{fig1}
\end{figure}

We show in particular that the coupling between two nonlinear nanomechanical resonators~(see Fig.~\ref{fig1}) realizes an effective \textit{phonon Josephson junction} exhibiting two different dynamical behaviors depending upon the strength of the mechanical nonlinearity: i) Josephson oscillation~(phonon-Rabi oscillation) at small nonlinearity; ii) self-trapping~(phonon blockade) at larger nonlinearity.
When the nonlinearity exceeds a critical value, a transition from a dynamical behavior to the other occurs and the exchange of excitations between the
two NMRs is suppressed. The effective Josephson oscillations between the two mechanical resonators are blocked and as a consequence, most phonons are self-trapped in one of two mechanical resonators.
The proposed scheme can be realized with nowadays technology~\cite{Croy}, and is suited to investigate a wide range of interesting phenomena such as the observation of spontaneous mirror-symmetry breaking~\cite{Hamel}, nonlinear phase dynamics and phase diffusion~\cite{Boukobza}, quantum chaos~\cite{Naether}, and phonon number squeezing in nonlinear nanomechanical resonators~\cite{Ferrini}.

The paper is organized as follows. After introducing the model in Sec. II, in Sec. III we find an effective Hamiltonian describing the phonon Josephson junction.  The dynamics of the system is studied in terms of the structure of its phase-space portrait. Then, we solve the equations for the time evolution of the phonon population imbalance, and we evaluate also the effects of the phonon decay. In Sec. IV, the Hamiltonian of the system will be represented in terms of angular momentum variables and the phonon-phonon coherence will be studied. Finally, concluding remarks are given in Sec. V.
\section{System}
We consider a system of two coupled nonlinear NMRs, such as two doubly clamped nanomechanical beams or nanotubes, and we restrict to study the dynamics of their fundamental flexural mode with frequency $ \omega_{0i } $, see Fig.~\ref{fig1}. The coupling between two NMRs can be realized by either mechanical or electrostatic/piezoelectric~\cite{Dykman, Karabalin, Kenig,Sotiris} coupling. The Hamiltonian describing the coupled
nonlinear resonators is~(see Appendix)
\begin{equation}\label{classicalham}
\hat H=\sum_{i=1,2}\Big[\frac{\hat{P}^2_i}{2m_i}+\frac{m_i \omega_{0i}^2\hat{X}^2_i}{2}\Big]+\sum_{i=1,2}\frac{\lambda_{0,i}}{4}\hat{X}_i^4-G_0 \hat{X}_1 \hat{X}_2,
\end{equation}
where~$\hat X_i$ and~$\hat P_i$~$(i=1,2)$ are the displacement and momentum operators of the~$i$th mechanical resonator with effective mass~$m_i$, while $ \lambda_{0i}\approx0.060\frac{m_i\omega_{0i}^2}{K_i^2} $ is the strength of nonlinearity with $ K_i $ being the ratio between the bending and compressional rigidities. The last term of Hamiltonian describes the coupling between two NMRs, with $ G_0 $ the inter-mode coupling rate.

By introducing the annihilation~$ \hat b_i $ and creation~$ \hat b_i^{\dagger} $ operators with commutators~$ [\hat b_i,\hat b_j^{\dagger}]=\delta_{ij} $, such that $ \hat X_i=\sqrt{\frac{\hbar}{2m_i\omega_{0i}}}(\hat b_i+\hat b_i^{\dagger}) $ and $ \hat P_i=i\sqrt{\frac{\hbar m_i\omega_{0i}}{2}}(\hat b_i^{\dagger}-\hat b_i) $, the Hamiltonian~(\ref{classicalham}) can be rewritten as
\begin{equation}\label{Hamiltonian0}
\hat H=\hbar\sum_{i=1}^{2}\Big[\omega_{0i} \hat b_i^{\dagger}\hat b_i+\frac{\tilde{\lambda}_i}{2}(\hat b_i^{\dagger}+\hat b_i)^4\Big]-\hbar \tilde{G}(\hat b_1^{\dagger}+\hat b_1)(\hat b_2^{\dagger}+\hat b_2),
\end{equation}
where $\tilde{\lambda}_i=\frac{\lambda_{0i}}{2}\frac{x_{0i}^4}{\hbar}  $, $\tilde{G}=\frac{G_0 x_{01}x_{02}}{\hbar} $, with $ x_{0i}=\sqrt{\frac{\hbar}{2m_i\omega_{0i}}} $ ($i=1,2$) being the mechanical zero-point motion amplitudes.

We move to the frame rotating at the resonance frequency of the first NMR, $\omega_{01}$, i.e., we perform the unitary transformation $ \hat U(t)=e^{-i\omega_{01}(\hat b_1^{\dagger}\hat b_1+\hat b^{\dagger}_2\hat b_2)t} $, and get the effective Hamiltonian $ \hat H_\mathrm{R}=\hat{U}^{\dagger}\hat H \hat U-i\hbar\hat{U}^{\dagger}\partial \hat U/\partial t $
\begin{eqnarray}\label{HamiltonianInt0}
\hat H_{\mathrm{R}}&=&\hbar\Delta_0 \hat{b}_2^{\dagger}\hat b_2+3\hbar\sum_{i=1}^{2}\tilde{\lambda}_i\Big[\hat b_i^{\dagger 2}\hat b_i^2+2\hat b_i^{\dagger}\hat b_i\Big] \nonumber\\
&-&\hbar G\Big[\hat b_1^{\dagger}\hat b_2^{\dagger}e^{i2\omega_{01} t}+\hat b_2\hat b_1e^{-i2\omega_{01} t}+\hat b_2^{\dagger}\hat b_1+\hat b_1^{\dagger}\hat b_2\Big]\nonumber\\
&+&\hbar\sum_{i=1}^{2}\tilde{\lambda}_i\Big[\hat b_i^{\dagger 4}e^{i4\omega_{01} t}+\hat b_i^4e^{-i4\omega_{01} t}\nonumber\\&+&e^{i2\omega_{01} t}(6\hat b_i^{\dagger 2}+4\hat b_i^{\dagger 3}\hat b_i)
+e^{-i2\omega_{01} t}(6\hat b_i+4\hat b_i^{\dagger}\hat b_i^3)\Big]
\nonumber,
\end{eqnarray}
where $\Delta_0 = \omega_{02}-\omega_{01}$. We then make the rotating wave approximation (RWA) and neglect the terms rotating at~$ \pm4\omega_{01} $ and~$ \pm2\omega_{01} $, which is justified when $ G_0,\lambda_i,\Delta_0 \ll \omega_{0i}$, so that the above Hamiltonian reduces to
\begin{equation}\label{HamiltonianInt}
\hat H_{\mathrm{R}}=\hbar \Delta_0 \hat{b}_2^{\dagger}\hat b_2+\hbar\sum_{i=1}^{2}\frac{\lambda_i}{2}\Big[2\hat b_i^{\dagger}\hat b_i+\hat b_i^{\dagger}\hat b_i^{\dagger}\hat b_i \hat b_i\Big]-\hbar G(\hat b_1^{\dagger}\hat b_2+\hat b_2^{\dagger}\hat b_1),
\end{equation}
where $ \lambda_i= \frac{3\lambda_{0i}x_{0i}^4}{\hbar} = 6 \tilde{\lambda}_i $ are the effective nonlinearity strength for NMRs, and~$ G= \frac{2G_0x_{01}x_{02}}{\hbar}= 2 \tilde{G}$ is the coupling rate between two NMRs.

\section{Phonon Josephson junction}
\subsection{Effective Hamiltonian}
We first neglect the effect of mechanical damping so that the dynamical behavior of the two interacting
nonlinear NMRs is described by the Heisenberg equations of motion associated with the Hamiltonian~(\ref{HamiltonianInt}), namely
\begin{subequations}\label{equationmotion}
\begin{eqnarray}
\dot{\hat b}_1&=&iG \hat b_2-i\lambda_1(1+\hat b_1^{\dagger}\hat b_1)\hat b_1,\\
\dot{\hat b}_2&=&i G \hat b_1-i\Big[
\Delta_0+\lambda_2(1+\hat b_2^{\dagger}\hat b_2)\Big]\hat b_2.
\end{eqnarray}
\end{subequations}
It is easy to verify that in this case the total phonon number~$N_T\equiv\hat n_1+\hat n_2 $ is a constant of motion. Typically one distinghishes between three different regimes according to the value of the dimensionless nonlinearity parameter~$g\equiv N_T(\lambda_1+\lambda_2)/4G$~\cite{Gati,Paraoanu}:~(i)~the quasilinear Rabi regime~$g< 1$;~(ii)~the intermediate Josephson regime $ 1<g<N_T^2 $, and (iii)~the quantum Fock regime~$g>N_T^2$. The linear Rabi regime corresponds to the strong-coupling regime where the mechanical nonlinearities are negligible compared to the coupling rate. This regime is well suited for coherent phonon manipulation~\cite{Okamoto,Yamaguchi,Weig}, and for coherent transfer of the phonon populations between the resonators, namely Rabi oscillations~\cite{Verhagen,Palomaki}. In the Josephson regime, the fluctuations of the phonon numbers are reduced but the coherence between mechanical resonators is strong~(see Appendix). Therefore, in this regime a relative phase $ \phi $ can be defined, which has only a small quantum mechanical uncertainty~($\Delta \phi\ll 1$). In this regime, for large phonon numbers~( i.e., $ N_T\gg1 $), the operators can be treated as classical quantities, $ \hat b_i\sim \sqrt{n_i}e^{i\theta_i} $ where $ n_i $ is the phonon population in $ i $th mechanical resonators whereas $ \theta_i $ is its phase. Finally, in the quantum Fock regime, the mechanical Josephson junction is dominated by the strong mechanical nonlinearities, thus the eigenstates have a well-defined phonon number in each resonators and as the coherence vanishes, the phase is completely undefined.

Here, we will not be interested in the quantum Fock regime and focus on the dynamics of the system in the first two, Rabi and Josephson, regimes. By introducing the fractional population imbalance, $ z(t)=\left[n_1(t)-n_2(t)\right]/N_T\in [-1,1] $, and relative phase $\phi(t)=\theta_2(t)-\theta_1(t)\in [0,2\pi]$,  Eqs.~(\ref{equationmotion}) reduce to
\begin{subequations}\label{classicalequation}
\begin{eqnarray}
\dot{z}(t)&=&-\sqrt{1-z^2(t)}\,\mathrm{sin}[\phi(t)],\\
\dot{\phi}(t)&=&\Delta + g z(t)+\frac{z(t)}{\sqrt{1-z^2(t)}}\,\mathrm{cos}[\phi(t)]
\end{eqnarray}
\end{subequations}
where $ \Delta=[-\Delta_0+(N_T/2+1)(\lambda_1-\lambda_2)]/2G $ and time has been rescaled so that $ 2G t\rightarrow t $. We notice that these equations are invariant under the transformation $ \phi\rightarrow -\phi+\pi$, $ \Delta \rightarrow -\Delta $ and $g\rightarrow -g $. We can now establish interesting analogies and differences with SJJ physics. In fact, we can view $ z $ and $ \phi $ as two classical conjugated variables, and the above equations as Hamilton equations derived from a classical effective Hamiltonian describing a phonon Josephson junction, $ \dot{z}=-\partial H_{J}/\partial \phi $ and  $ \dot{\phi}=\partial H_{J}/\partial z $, with
\begin{equation}\label{PJJhamm}
H_{J}=\Delta z +\frac{g}{2}z^2-\sqrt{1-z^2}\,\mathrm{cos}\,\phi.
\end{equation}
The above Hamiltonian is similar to the Hamiltonian of a SJJ but it differs in its nonlinearity in $ z $ as in SJJ~(considering two equal-volume superconducting grains) the charge leakage through the external circuit strongly suppresses population imbalances, i.e., $ n_1\simeq n_2 \simeq N_T/2 $ and $ z \simeq 0 $~\cite{Barone,Ohta}. In this case, in the resistively and capacitively
shunted junction model, the SJJ is analogous to a rigid pendulum, while Hamiltonian~(\ref{PJJhamm}) becomes analogous to a nonrigid, momentum-shortened, pendulum of length $ \sqrt{1-z^2} $ and tilt angle $ \phi $~\cite{Smerzi,Raghavan}. Nonetheless, we can maintain a close connection with the SJJ physics and, in analogy with the Cooper-pair tunneling current in SJJ, we can define an effective phonon tunneling current $ I=\frac{N_T}{2}\dot{z}=I_c\sqrt{1-z^2}\,\mathrm{sin}\phi $, where $ I_c=G N_T $.

\subsection{Fixed-Energy Trajectories}
The Hamiltonian of Eq.~(\ref{PJJhamm}) describes a system with one degree-of-freedom and therefore an integrable dynamics with no chaos. As a consequence, the phase-space trajectories of the system follow the lines of constant (conserved) energy.
The basic structure of phase-space trajectories can be determined by finding the stationary points of the dynamics, given by setting $ \partial H_{J}/\partial \phi=0 $ and  $ \partial H_{J}/\partial z=0 $. The first equation provides two possible set of stationary values for the relative phase: zero phase, $ \phi_s=2n\pi $, and $ \pi $-phase, $ \phi_s=(2n+1)\pi $, (with integer $ n$). Substituting these values of $ \phi_s $ into the second equation yields two equations for the stationary value $z_s$, i.e. $z_s(g+\frac{1}{\sqrt{1-z_s^2}})+\Delta=0$ and
$z_s(g-\frac{1}{\sqrt{1-z_s^2}})+\Delta=0$, respectively. Restricting to the case $ \Delta=0 $, the minimum energy stable steady state of the system is given by the zero phase solution $ [\phi_s=2n\pi, z_s=0 ]$, with energy $ \mathcal{E}=-1 $, while the unstable steady state depends upon the value of the nonlinearity parameter $g$: it is given by  $ [\phi_s=(2n+1)\pi, z_s=0 ]$ with energy $ \mathcal{E}=1 $ with $ [\phi_s=(2n+1)\pi, z_s=0 ]$ when $g < 1$, but one has two degenerate energy maxima at
$[\phi_s =(2n+1)\pi,\, z_s=\pm \sqrt{1-g^{-2}}] $ with~$ \mathcal{E}=\frac{g}{2}(1+g^{-2}) $ when $g>1$.

\begin{figure}[t]
\centering
\includegraphics[width=3in]{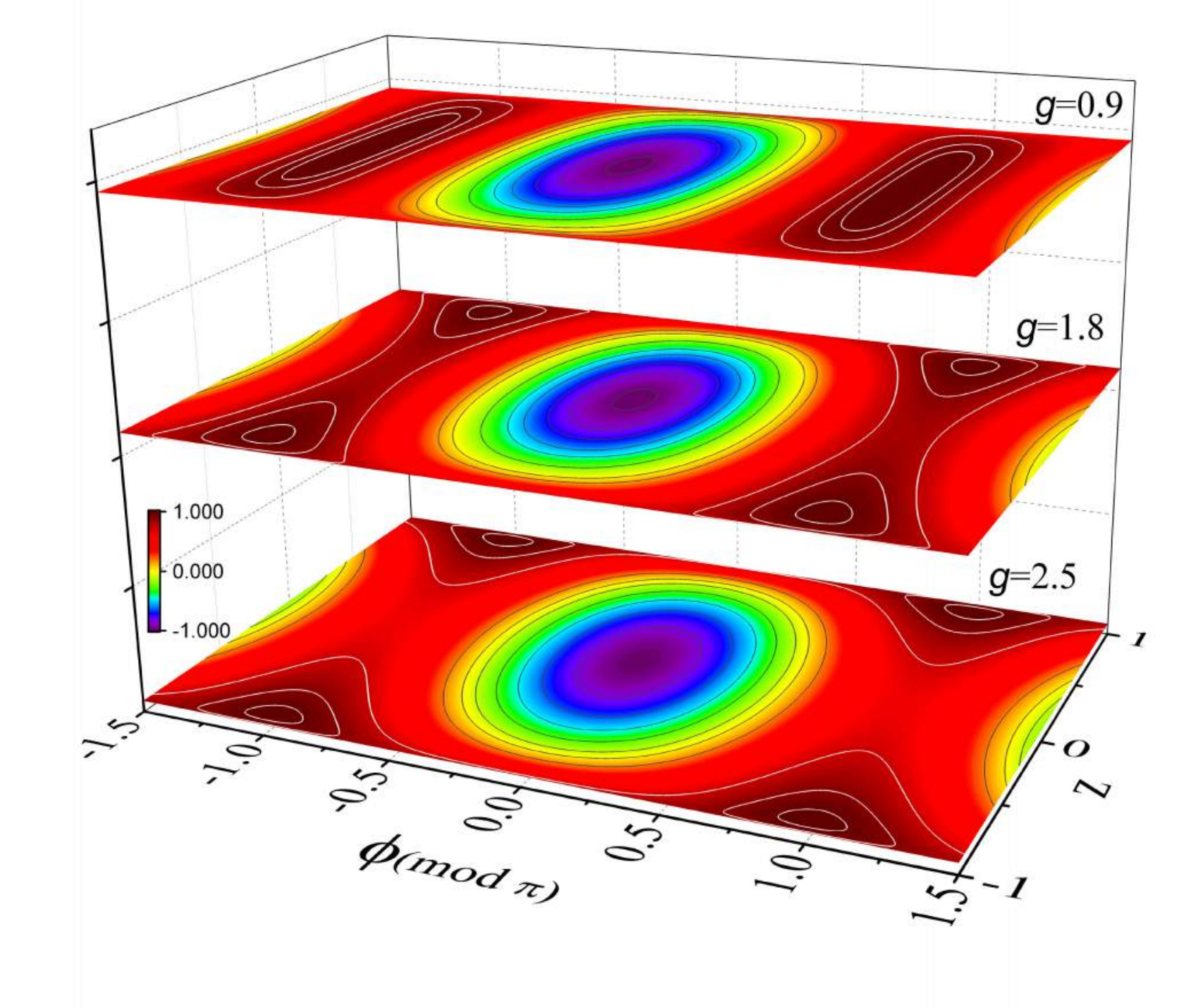}
\caption{Energy contours of the phonon Josephson junction versus $ z $ and $\phi$ at $\Delta =0$, for: $ g=0.9 $,~$ g=1.8$, and ~$g=2.5 $.}
\label{contour}
\end{figure}

Figure~\ref{contour} shows the energy contours of a phonon Josephson junction for different values of parameter $ g $.  It is evident that the location of the energy minima, maxima and saddle points crucially depends upon the dimensionless parameter $ g $. For $ g\leq 1$~(strong coupling), the minima are at
$ [z,\phi]=[0,2n\pi] $ and the maxima settle in $  [z,\phi]=[0,(2n+1)\pi] $, whereas for $ g>1$~(strong nonlinearities), the minima are still at $  [z,\phi]=[0,2n\pi] $, while $ [z,\phi]=[0,(2n+1)\pi] $ becomes saddle points, and maxima move to the two new locations~$[z,\phi]=\big[\pm\sqrt{1-g^{-2}},(2n+1)\pi\big]$. This transition of the point $ [z,\phi]=[0,(2n+1)\pi] $ from a local maximum to
a saddle point, and the appearance of two new maxima for $ g>1$, are a manifestation of the existence of a running-phase and of $ \pi $-phase self-trapping states.

\subsection{System dynamics}
\begin{figure}[t]
\centering
\includegraphics[width=3.5in]{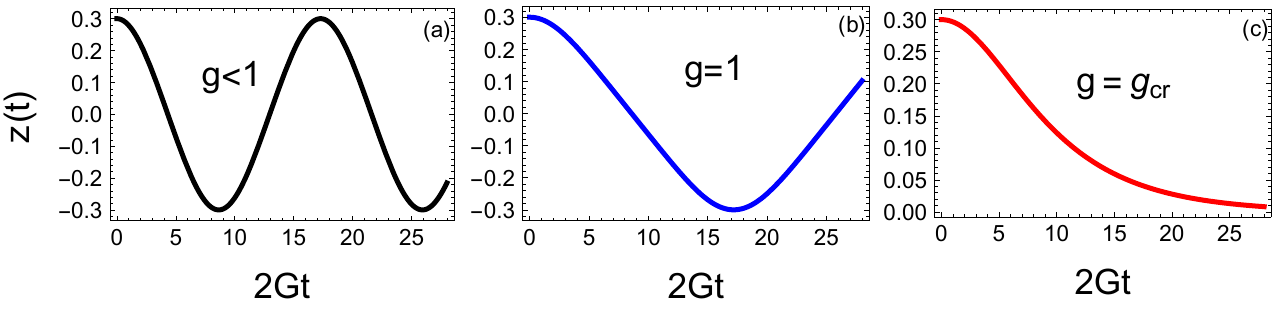}
\includegraphics[width=3.5in]{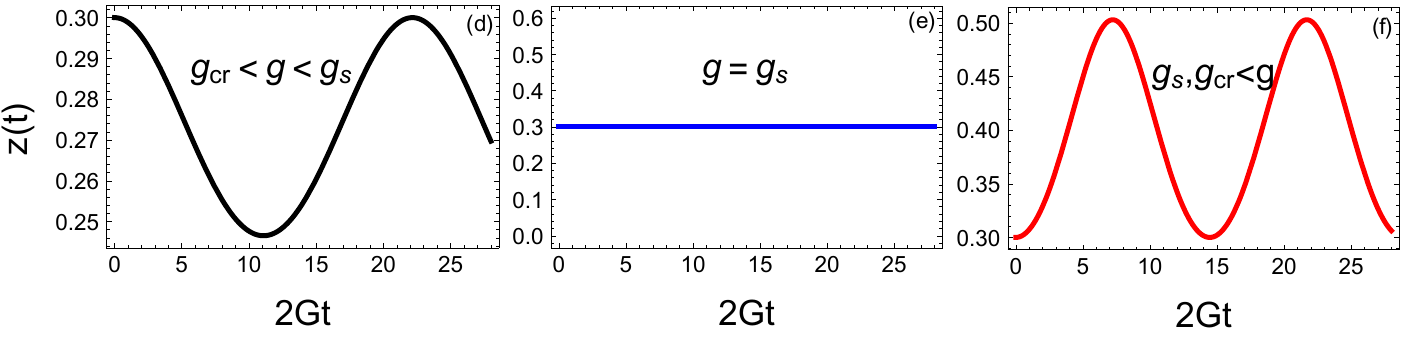}
\caption{Phonon population imbalance~$ z(t) $ as a function of the rescaled time~$ 2G t $ when $\Delta =0$ and for the specific choice of initial condition $ [z(0),\phi(0)]=[0.3,\pi] $. We consider different values of the parameter $ g $:~(a)$ g=0.9 $, (b)~$ g=1 $,(c)~$ g=g_{cr}\simeq 1.02357 $,~(d)~$ g=1.04 $ where $ g_{\mathrm{cr}}<g<g_s $, (e)~$g=g_s\simeq1.04828$, and (f)~$ g_{\mathrm{cr}},g_s<g=1.056 $. }
\label{imbalance}
\end{figure}

The dynamics of the system is obtained by solving Eqs.~(\ref{classicalequation}). We restrict again to the case $\Delta =0$ which is the more interesting one, corresponding to the limit of identical MRs. Figure~\ref{imbalance} describes the time evolution of population imbalance $ z(t) $ versus rescaled time~$ 2G t $, for different values of parameter $ g $, showing the transition from the Rabi oscillation to the Josephson and self-trapping regime, for the specific choice of initial conditions $ [z(0),\phi(0)]=[0.3,\pi] $.

In the limit of very small nonlinearity, $ g\ll 1$, Eqs.~(\ref{classicalequation}) become
$\dot{z}(t)\simeq-\phi(t) $ and $\dot{\phi}(t)\simeq(g -1)z(t)$, which describe the small-amplitude oscillations of a pendulum in the $ \pi $-phase mode with a frequency~$\omega_{\pi}=2G\sqrt{1-g} \simeq 2G$. This corresponds to harmonic Rabi-like oscillations in the phonon population of each mechanical resonator with the same frequency~\cite{Okamoto,Yamaguchi} (see Fig.~\ref{imbalance}(a)). As the parameter $ g $ increases, the oscillations become anharmonic and the system moves into the Josephson regime (see Fig.~\ref{imbalance}(b)).
The dynamics change significantly above a critical value of the effective nonlinearity $g \geq g_{\mathrm{cr}}$, as shown in Figs.~\ref{imbalance}(c)-(f): the phonon number in each mechanical resonator oscillates around a \emph{nonzero} time-averaged population imbalance, $ \langle z(t)\rangle\neq 0 $, meaning that the phonon populations become macroscopically self-trapped~(MST). The critical value $g_{\mathrm{cr}}$ depends upon the explicit value of the initial conditions $ [z(0),\phi(0)]$  and is related to the condition that the corresponding initial classical energy is larger than that of the saddle point $  [z,\phi]=[0,(2n+1)\pi] $, $H_J[z(0),\phi(0)]=\frac{g}{2}z(0)^2-\sqrt{1-z^2}\,\mathrm{cos}[\phi(0)]>1 $, yielding the critical parameter for MST $$
g_{\mathrm{cr}}=\frac{1+\sqrt{1-z^2(0)}\,\mathrm{cos}[\phi(0)]}{z^2(0)/2}. $$
In Figs~\ref{imbalance}(d)-(f) the time-averaged value of the phase is $ \langle \phi \rangle =\pi $: these modes, known as $ \pi $-phase modes, describe the effective tunneling of the phononic excitation from one MR to the other. The numerical solution in Figs.~\ref{imbalance}(d)-(f) shows also that in the MST regime $g>g_{\mathrm{cr}}$ ($g_{\mathrm{cr}}\sim 1.02357$ when $ [z(0),\phi(0)]=[0.3,\pi] $ as in Fig.~\ref{imbalance}), one can observe two different types of $ \pi $-phase modes: (i) when the time-averaged population imbalance is smaller than the unstable stationary value $|z_s|=\sqrt{1-g^{-2}}$, $ \langle z(t)\rangle < |z_s| $; (ii) when $ \langle z(t)\rangle> |z_s| $. Which kind of self-trapping occurs depends again upon the value of $g$: the first MST mode with $ \langle z(t)\rangle < |z_s| $ occurs when $g_{\mathrm{cr}}<g<g_s$ with $ g_s=1/\sqrt{1-z^2(0)}$ ($g_s\sim 1.04828 $ when $ [z(0),\phi(0)]=[0.3,\pi] $ as in Fig.~\ref{imbalance}(e)). Instead the second MST mode with $ \langle z(t)\rangle > |z_s| $ occurs when $g>g_s$ (see Fig.~\ref{imbalance}(f)).
When $ g=g_s $ (see Fig. \ref{imbalance}(e)), the system settles in an intermediate regime where there is no oscillation of the population imbalance.

\subsection{Effects of phonon decay}
We now include the effect of mechanical damping, which can be described by adding phonon loss terms
to the Heisenberg equations of Eq.~(\ref{equationmotion}). The resulting equations for both MRs become
\begin{subequations}\label{equationmotion2}
\begin{eqnarray}
\dot{\hat b}_1&=&i\Big[G \hat b_2-\lambda_1 (1+\hat b_1^{\dagger}\hat b_1)\hat b_1\Big]-\frac{\kappa_1}{2} \hat b_1,\\
\dot{\hat b}_2&=&i\Big[G \hat b_1-\Delta_0\hat b_2+\lambda_2(1+\hat b_2^{\dagger}\hat b_2)\hat b_2\Big]-\frac{\kappa_2}{2} \hat b_2,
\end{eqnarray}
\end{subequations}
where~$ \kappa_{i} $ are the MRs damping rates. For simplicity, we assume that the two NMRs have the same
loss rates, i.e~$ \kappa_0=\kappa_1=\kappa_2 $. We can again adopt the semiclassical approach of the previous subsection and define $ b_i=\langle \hat b_i\rangle=|b_i|e^{i\theta_i} $ with $ n_i=|b_i|^2 $ being the phonon number in \textit{i}th MR, and define again the quantities $z(t)$ and $\phi(t)$. In this case the total phonon number is no more conserved and we have to add a new dynamical variable associated with phonon loss, $ N=\langle \hat N \rangle $ with $ \hat N=\frac{ \hat b^{\dagger}_1 \hat b_1+\hat  b^{\dagger}_2 \hat b_2}{N_T} $, where $ N_T $ is the total phonon number at time $ t=0 $. The evolution equations for these parameters can be calculated from Eqs.~(\ref{equationmotion2}), getting
\begin{subequations}\label{classicalequation2}
\begin{eqnarray}
\dot{z}(t)&=&-\sqrt{N^2(t)-z^2(t)}\,\mathrm{sin}[\phi(t)]-\kappa z(t),\\
\dot{\phi}(t)&=&\Delta_{\kappa}+g\,z(t)+\frac{z(t)}{\sqrt{N^2(t)-z^2(t)}}\,\mathrm{cos}[\phi(t)],\\
\dot{N}(t)&=&-\kappa N(t),
\end{eqnarray}
\end{subequations}
where $ \Delta_{\kappa}=[-\Delta_0+(N_T e^{-\kappa t}/2+1)(\lambda_1-\lambda_2)]/2G $, $ \kappa=\kappa_0/(2G) $ and we have again rescaled the time so that $ 2G t\rightarrow t $. As expected, when phonon losses are negligible, $ \kappa_0\ll G \Leftrightarrow \kappa \sim 0$, $ N\simeq 1 $, and the above equations reduce to the simple forms presented in Eq.~(\ref{classicalequation}).

\begin{figure}[ht]
\centering
\includegraphics[width=3.5in]{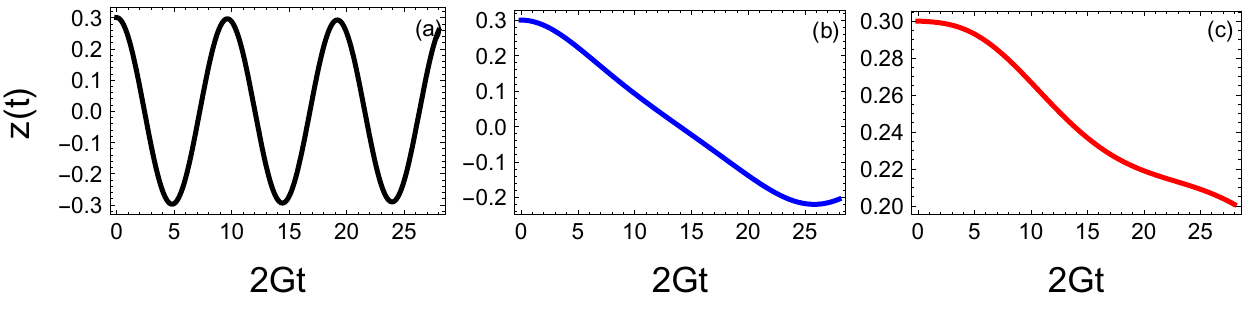}
\caption{Phonon population imbalance~$ z(t) $ as a function of rescaled time~$ 2G t $, when $\Delta =0$, for the specific choice of initial condition $ [z(0),\phi(0)]=[0.3,\pi] $, and in presence of phonon loss, $\kappa=0.001$. We consider different values of the parameter $ g $:~(a)~$ g=0.9 $, (b)~$ g=g_{\mathrm{cr}}\simeq 1.023 $, and ~(c)~$g=g_s=1.04828$. In all cases $ z(t) \to 0$ at long times due to the effect of mechanical damping.}
\label{damping}
\end{figure}

Fig.~\ref{damping} shows the time evolution of the phonon population imbalance~$ z(t) $ as a function of rescaled time~$ 2G t $ in the presence of small but nonzero phonon loss. This figure shows that the transient dynamics is similar to the one without damping shown in the previous subsection, i.e., phonons shuttle between the two MRs. However phonon losses significantly change the dynamics at long time scales because the population imbalance always tends to zero at long times, even at large nonlinearities $g$, leading to the suppression of phonon self-trapping.
\section{Phonon-Phonon Coherence}

In the previous Sections we have analyzed the phonon population dynamics in a semiclassical regime for the two NMRs. We now come back to the quantum nonlinear Hamiltonian of Eq.~(\ref{HamiltonianInt}) obtained after taking the RWA, in order to understand if and when the coupling allows to establish quantum coherence between the two NMRs. For this purpose it is convenient to use the representation of two bosonic systems in terms of angular momentum operators ~\cite{GMilburn,Zoller,Trimborn}, and define $\hat{J}_x=(\hat b_1^{\dagger}\hat b_2+\hat b_2^{\dagger}\hat b_1)/2$ the phonon tunneling operator, $\hat{J}_y=-i(\hat b_1^{\dagger}\hat b_2-\hat b_2^{\dagger}\hat b_1)/2 $ the current operator, and $\hat{J}_z\equiv\hat{n}=(\hat b_1^{\dagger}\hat b_1-\hat b_2^{\dagger}\hat b_2)/2  $ the number imbalance operator. The Casimir invariant $\hat{J}^2=\hat{J}_x^2+\hat{J}_y^2+\hat{J}_z^2$ is a function of the total phonon number operator $\hat{N}_T =\hat b_1^{\dagger}\hat b_1+\hat b_2^{\dagger}\hat b_2$, i.e., $\hat{J}^2=(\hat{N}_T/2)(\hat{N}_T/2+1)$, and it is therefore a constant of motion in the absence of phonon losses, which we will assume again in this Section. Therefore, within the subspace at fixed total phonon number $\hat{N}_T =N_T$, the state of the coupled NMRs system can be described in terms of the angular momentum vector $\vec{\textbf{J}}=(\hat J_x,\hat J_y,\hat J_z) $ with fixed modulus $J=N_T/2$, and the dynamics is driven by the Hamiltonian of Eq.~(\ref{HamiltonianInt}), which can be rewritten, modulo an irrelevant constant depending upon $N_T$, as
\begin{equation}\label{spinHamiltonian}
\hat H_{\mathrm{int}}=\hbar\Big(\frac{\lambda_1+\lambda_2}{2}\Big)\Big(\hat{J}_z-n_0\Big)^2-2\hbar G\hat{J}_x.
\end{equation}
Here, $ n_0 $ is similar to the parameter $\Delta$ of the semiclassical equation, and it is related to the eventual asymmetry between the two NMRs,
\begin{equation}
n_0=\frac{(\lambda_2-\lambda_1)(N_T+1)/2+\Delta_0}{\lambda_1+\lambda_2}.
\end{equation}
We restrict again to the Rabi and Josephson regime of not too large nonlinearities, $(\lambda_1+\lambda_2)\ll N_T G  $; the Hamiltonian of Eq.~(\ref{spinHamiltonian}) suggests that in this limit the ground state of the system is very close to the eigenstate with maximum $\hat{J}_x$, i.e., $\hat{J}_x=N_T/2$, with an energy eigenvalue close to $-\hbar GN_T$. Such a state however is just an example of \emph{spin coherent state}~\cite{arecchi}, which can be seen as semiclassical states of a spin which can be represented on the Bloch sphere as a disk of diameter
$  \sqrt{N_T}/2$, centered around the expectation value of the angular momentum operator with orientation $(\theta,\phi)$, $\langle\theta,\phi|\vec{\textbf{J}}|\theta,\phi \rangle=N_T(\mathrm{sin}\,\theta\,\mathrm{cos}\,\phi,~\mathrm{sin}\,\theta~\mathrm{sin}\,\phi,\,-\,\mathrm{cos}\,\theta)/2 $. A spin coherent state is defined as
\begin{equation}
|\theta,\phi \rangle=\sum_{m=-N_T/2}^{N_T/2} {N_T \choose m+N_T/2}^{1/2}\frac{\alpha^{m+N_T/2}}{(1+|\alpha|^2)^{N_T/2}}|m\rangle,
\end{equation}
with $ \alpha\equiv \mathrm{tan}(\theta/2)\mathrm{exp}(-i\phi) $ and $|m\rangle$ denotes the eigenstates of $\hat{J}_z$, $\hat{J}_z |m\rangle =m|m\rangle$. Therefore in the limit $(\lambda_1+\lambda_2)\ll N_T G  $ the ground state of the system of coupled NMRs is the spin coherent state $|\pi/2,0\rangle$ and we expect that in its time evolution, the quantum state of the system can be satisfactorily approximated by a time-dependent spin coherent state.

These spin coherent states at the same time represent states with large coherence between the two NMRs. In fact, it is easy to verify that one can rewrite $\hat{J}_x=(\hat b_+^{\dagger}\hat b_+-\hat b_-^{\dagger}\hat b_-)/2  $, with $\hat b_{\pm}=(\hat b_1\pm \hat b_2)/\sqrt{2}$, and therefore the approximate ground state $|\pi/2,0\rangle$ is a state with $N_T$ phonons in mode $\hat b_+$ and no phonon in the orthogonal mode $\hat b_-$. More in general, it is easy to show that a spin coherent state can be rewritten as
\begin{equation}
|\theta,\phi \rangle \propto \Big[\mathrm{exp}(-i\phi)\,\mathrm{sin}\,(\theta/2)\hat b_1^{\dagger}+\mathrm{cos}\,(\theta/2)\hat b_2^{\dagger}\Big]^{N_T}|0\rangle,
\end{equation}
where $ |0\rangle $ is the vacuum state with no phonons in the two NMRs. This means that in a spin coherent state, the two NMRs have perfectly locked phases and they share $N_T$ phonons which all occupy the same effective one-phonon mode.

\begin{figure}[ht]
\centering
\includegraphics[width=3.5in]{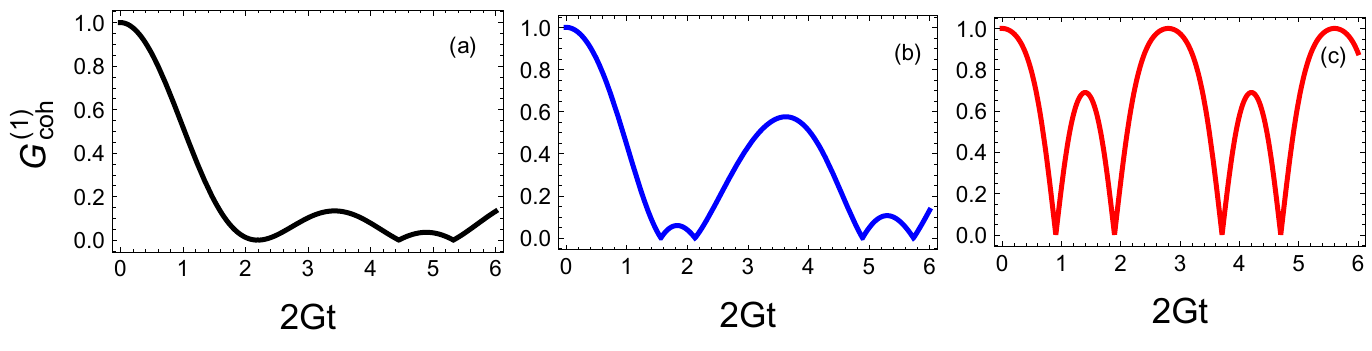}
\includegraphics[width=3.5in]{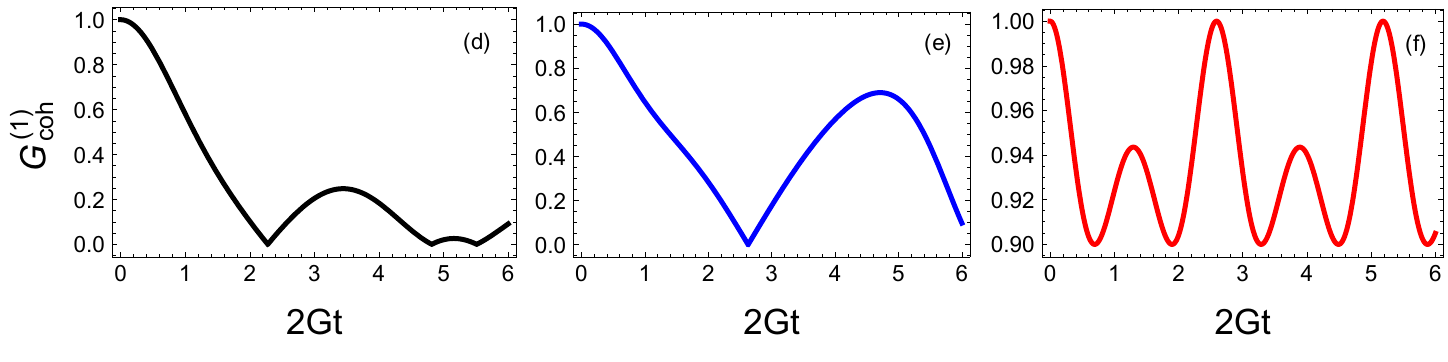}
\caption{Fringe visibility $ G^{(1)}_{coh}=\frac{2|\langle \hat J_x\rangle|}{N_T} $ versus rescaled time $ 2Gt $. The nonlinarity parameter is $ g=0.3 $~(a),(d), $ 1 $~(b),(e), and $  6 $~(c),(f). The system is starting from the spin coherent states $|\pi/2,\pi \rangle $ (a)-(c) and $|\pi/2,0 \rangle $ (d)-(f).}
\label{fringe}
\end{figure}

As mentioned above, in the Josephson regime $(\lambda_1+\lambda_2)\ll N_T G  $  we expect a semiclassical dynamics where the effective spin $\vec{\textbf{J}}=(\hat J_x,\hat J_y,\hat J_z) $ behaves like a classical quantity. Under this condition we replace the spin operator  $\vec{\textbf{J}} $ by $ c $ numbers, and exploit the factorization of expectation values of products of operators like, e.g., $ \langle \lbrace \hat J_x ,\hat J_y \rbrace \rangle$ by $2\langle \hat J_x \rangle \langle\hat J_y \rangle$. Starting from the Heisenberg equations for $\vec{\textbf{J}} $, one obtains by using a semiclassical limit that the angular coordinates~$ (\theta,\phi) $ satisfy
\begin{subequations}\label{stateequation}
\begin{eqnarray}
\dot{\theta}&=&-\mathrm{sin}(\phi),\\
\dot{\phi}&=&-g\,\mathrm{cos}(\theta)-\mathrm{cot}(\theta)\,\mathrm{cos}(\phi),
\end{eqnarray}
\end{subequations}
where the time has been rescaled $ 2G t\rightarrow t $.
However, we expect that the small but nonzero nonlinearity will affect such evolution and perturb the spin coherent states. This would manifest in a loss of coherence between the two NMRs, due to the interplay between the mechanical coupling (phonon tunneling) and the nonlinearity. In fact, due to tunneling, different $  J_z $ eigenstates oscillate with different frequencies, and the evolution under the
Hamiltonian~(\ref{spinHamiltonian}) leads to a loss of
phonon-phonon coherence, which can be quantified by the mean fringe visibility~\cite{Wineland},
\begin{equation}
G^{(1)}_{coh}=\frac{2|\langle \hat J_x\rangle|}{N_T}.
\end{equation}
By solving the coupled equations~(\ref{stateequation}) we can evaluate the the mean fringe visibility $ G^{(1)}_{coh}=\mathrm{sin}\,\theta\,\mathrm{cos}\,\phi$.
Here, we study the~$ g $ dependence of the fringe-visibility evolution: in Figs.~\ref{fringe}(a)-(b) we plot the numerically calculated $ G^{(1)}_{coh}$ for three values of parameter $ g $ when the initial state is the excited coherent state $ |\pi/2,\pi \rangle $. For the $ g=0.3 $, see Fig.~\ref{fringe}(a),  phase locking is not attained and the phonon coherence decays quickly. However, by increasing the parameter $ g $, see Fig.~\ref{fringe}(b)-(c), the expected phase locking is obtained even by a very weak coupling, leading
to appearance of collapse and revival of coherence in the fringe visibility. Moreover, Figs.~\ref{fringe}(d)-(f) show a similar situation with a different initial state $ |\pi/2,0 \rangle $~(ground state). It is evident that for small values of parameter $ g $ the system shows collapse and revival in the fringe visibility, whereas for increasing parameter $ g $, one gets a nonvanishing value of the fringe visibility, and therefore a nonvanishing phonon coherence is maintained in time.

\section{Summary and conclusions}
In summary, we have shown that the coupling between two nonlinear nanomechanical resonators realizes an interesting analog of a Josephson junction. The dynamical behavior of the system has been studied in two different regimes: Josephson oscillation~(phonon-Rabi oscillation) and macroscopic self-trapping~(phonon blockade). We have shown that when the mechanical nonlinearities is larger than a critical value, the phonon-Josephson oscillation between the two mechanical resonators is completely blocked and phonons are self-trapped. Moreover, an effective classical Hamiltonian for the phonon Josephson junction has been derived and its mean-field dynamics has been studied in phase space. Finally, we has studied the phonon-phonon coherence quantified by the mean fringe visibility, and we have shown that the mechanical coupling leads to a loss of coherence between the two mechanical resonators.

The dynamics of the phonon population, and therefore Josephson oscillations and self-trapping phenomena, could be experimentally verified in a resolved sideband optomechanical setup, using the phonon counting techniques recently demonstrated in Ref.~\cite{Painter2}. The detection of mechanical fringe visibility and phonon coherence instead requires measuring the correlations between the two phonon fields. This correlation measurement could be in principle realized by first transferring the phonon states onto optical fields using an additional optomechanical coupling, as first suggested in Ref.~\cite{vitali} and then realized e.g. in Ref.~\cite{Palomaki}, and then measuring the corresponding optical correlations with an heterodyne technique.

Our scheme could potentially be used for the observation of spontaneous mirror-symmetry breaking~\cite{Hamel}, or for studying nonlinear phase dynamics, phase diffusion~\cite{Boukobza}, quantum chaos~\cite{Naether}, and phonon number squeezing~\cite{Ferrini} in coupled nonlinear mechanical resonators.

\section*{acknowledgement}
The authors would like to thank E. Weig, J. Fink, P. Rabl, D. P. DiVincenzo, M. Goldsche, T. Khodkov, G. Verbiest, and C. Stampfer, for valuable discussions. The work of S. B. has been supported by the European Commission via the SCALEQIT program and by the Alexander von Humboldt Foundation. The work of D. V. has been supported by the European Commission via the ITN-Marie Curie project cQOM and the FET-Open Project iQUOEMS.

\appendix
\section*{Appendix}
In this Appendix we provide the detailed derivation of the starting Hamiltonian of the scheme, Eq.~(\ref{classicalham}), exploiting the theory of elasticity of a nonlinear mechanical resonator and also how one can introduce the coupling between the two nanomechanical resonators.
\section{Elasticity-Theory of a Nonlinear Mechanical Resonator}
\begin{figure}[h]
\centering
\includegraphics[width=2.5in]{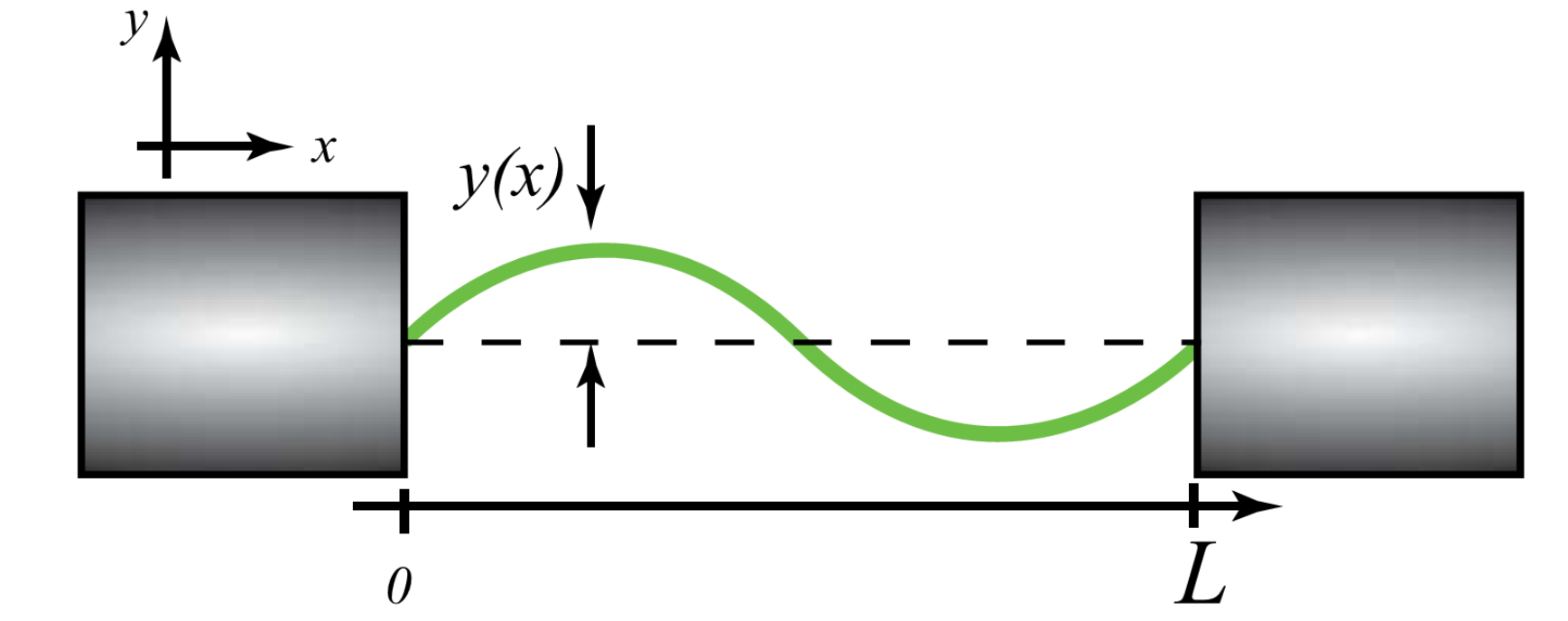}
\caption{~Schematic description of a doubly clamped nanomechanical resonator with length $ L $. }
 \label{figbeam}
\end{figure}

We consider a doubly clamped mechanical resonator~(MR) with a constant linear mass density $ \mu $ and length~$ L $  in which the cross-section of the beam is much smaller than its length (see Fig.~\ref{figbeam}). We also assume that the planar deflection in the transverse direction of the beam is described by $ y(x) $ for $0 \leq x\leq L$. The Lagrangian of the MR can be written on the basis of the theory of elasticity of a beam~\cite{Landau, Rips1,cleland}
\begin{equation}\label{lag}
\mathcal{L}[y(x)]=\frac{\mu}{2}\int dx \frac{d y(x)}{dt}-V_{B}(x),
\end{equation}
where
\begin{equation}
V_{B}(x)=\frac{I}{2}\int dx   \Big[K\frac{d^2 y(x)}{dx^2}\Big]^2
\end{equation}
describes the MR bending energy, $ I=E A $ is the linear modulus of a MR rod with cross-sectional area $ A $ and Young modulus $ E $. The parameter $ K $ is the ratio between the bending and compressional rigidities, whose value depends upon the cross-sectional geometry of the doubly clamped MR: \textit{i})~for a rectangular cross section of thickness $ d $, $ K=d/\sqrt{12} $; \textit{ii})~for a circular cross section with radius $ R $, $ K=R/2 $; \textit{iii})~for a cylindrical shell~(such as a nanotube) $ K=R/\sqrt{2} $. We consider a two-sided clamped MR where the end points at $ x=0 $ and $ x=L $ are fixed, which means that $ y(0)=y(L)=0 $ and $y'(0)=y'(L)=0 $. In the absence of dissipation and other external forces, the dynamical behavior of the flexural mode is given by the Lagrangian of Eq.~(\ref{lag}), leading to following equation of motion
\begin{equation}\label{equationmotionbis}
\mu \frac{\partial^2 y}{\partial t^2}+K^2 I \frac{\partial^4 y}{\partial x^4}=0.
\end{equation}
The eigenmodes of this equation are
\begin{eqnarray}\label{eigenmode}
\psi_n(x)&=&\frac{1}{M_n}\Big[\frac{\mathrm{sin}(\zeta_n x/L)-\mathrm{sinh}(\zeta_n x/L)}{\mathrm{sin}(\zeta_n)-\mathrm{sinh}(\zeta_n)}
\nonumber\\
&-&\frac{\mathrm{cos}(\zeta_n x/L)-\mathrm{cosh}(\zeta_n x/L)}{\mathrm{cos}(\zeta_n)-\mathrm{cosh}(\zeta_n)}\Big],
\end{eqnarray}
where the eigenvalues $ \zeta_n $ satisfies the transcendental equation $ \mathrm{cos}(\zeta_n)\mathrm{cosh}(\zeta_n)=1 $, with solutions $ \zeta_n=4.73, 7.85,... $ . The parameters $ M_n $ represents the normalization constants chosen such that $\mathrm{max}\{\psi_n(x)\}=1 $. Using the standard Legendre transformation for deriving the Hamiltonian from the Lagrangian, and expanding the beam's deflection in term of its eigenmodes,
\begin{equation}\label{solution}
y(x,t)=\sum_n \psi_n(x)X_n(t),
\end{equation}
we finally get the beam's Hamiltonian
\begin{equation}\label{hamiltonianH0}
H_0=\sum_n \Big(\frac{P_n^2}{2m_n}+\frac{1}{2}m_n\Omega_n^2X_n^2\Big),
\end{equation}
where  $ m_n=\mu \int_0^L \psi^2_n(x)dx$ and $ \Omega_n=\sqrt{\frac{I K^2}{\mu}}(\frac{\zeta_n}{L})^2  $ are respectively the effective mode mass and the vibrational frequency of the $ n $th mode with the deflection $ X_n $ and momentum $ P_n=m_n\frac{\partial X_n}{\partial t} $.

The harmonic Hamiltonian (\ref{hamiltonianH0}) does not fully describe the whole energy of the clamped beam. We need to add a correction term originating from a stretching effect that occurs due to the deflection if the end points of the MR are fixed~\cite{Rips1,carr}. The stretching energy of the beam is described by
\begin{equation}
V_E=\frac{I}{2L}(L_t-L)^2\simeq \frac{I}{8L}\Big[\int dx (\frac{d y}{dx})^2 \Big]^2,
\end{equation}
where the total stretched length is $ L_t=\int dx \sqrt{1+(\frac{d y}{dx})^2}\simeq L+\frac{1}{2} \int dx (\frac{d y}{dx})^2 $ , with $ L $ being the zero deflection length. Including the the elastic potential $ V_E $ into the Hamiltonian of Eq.~(\ref{hamiltonianH0}), we obtain a nonlinear Hamiltonian for the MR,
\begin{equation}\label{Hamiltonianmulti}
H_{r}=\sum_n \Big(\frac{P_n^2}{2m_n}+\frac{m_n\Omega_n^2}{2}X_n^2\Big)+\frac{I}{8L}\sum_{i,j,k,l}(N_{ij}N_{kl})X_iX_jX_kX_l,
\end{equation}
where $ N_{ij}=\int_0^L \psi'_i(x)\psi'_j(x)dx $.

The Hamiltonian in Eq.~(\ref{Hamiltonianmulti}) describes the multimode and nonlinear Hamiltonian for a doubly clamped MR. For the sake of simplicity, we can restrict our analyzes to the fundamental mode $ n=1 $ only, because the terms involving higher order modes induce smaller frequency shifts and smaller nonlinearity. Therefore, the Hamiltonian in Eq.~(\ref{Hamiltonianmulti}) reduces to the standard Hamiltonian of the Duffing oscillator
\begin{equation}\label{Hamiltoniansingle}
H_{r}=\frac{P^2}{2m}+\frac{1}{2}m\omega_0^2X^2+\lambda_0 X^4,
\end{equation}
where $ \omega_0\equiv \Omega_1 $ is the fundamental frequency, $ m\equiv m_1\simeq 0.3965 \mu L $ is the effective mass of the fundamental mode, and
\begin{equation}
\lambda_0=\frac{N_{11}^2 L^3\mu  \omega_{0}^2}{2 \zeta_1^4K^2}\approx 0.060 m \frac{\omega_0^2}{K^2}.
\end{equation}
represents the mechanical anharmonicity.

\section{Coupling Between Two Nanomechanical Resonators}
\begin{figure}[ht]
\centering
\includegraphics[width=2.5in]{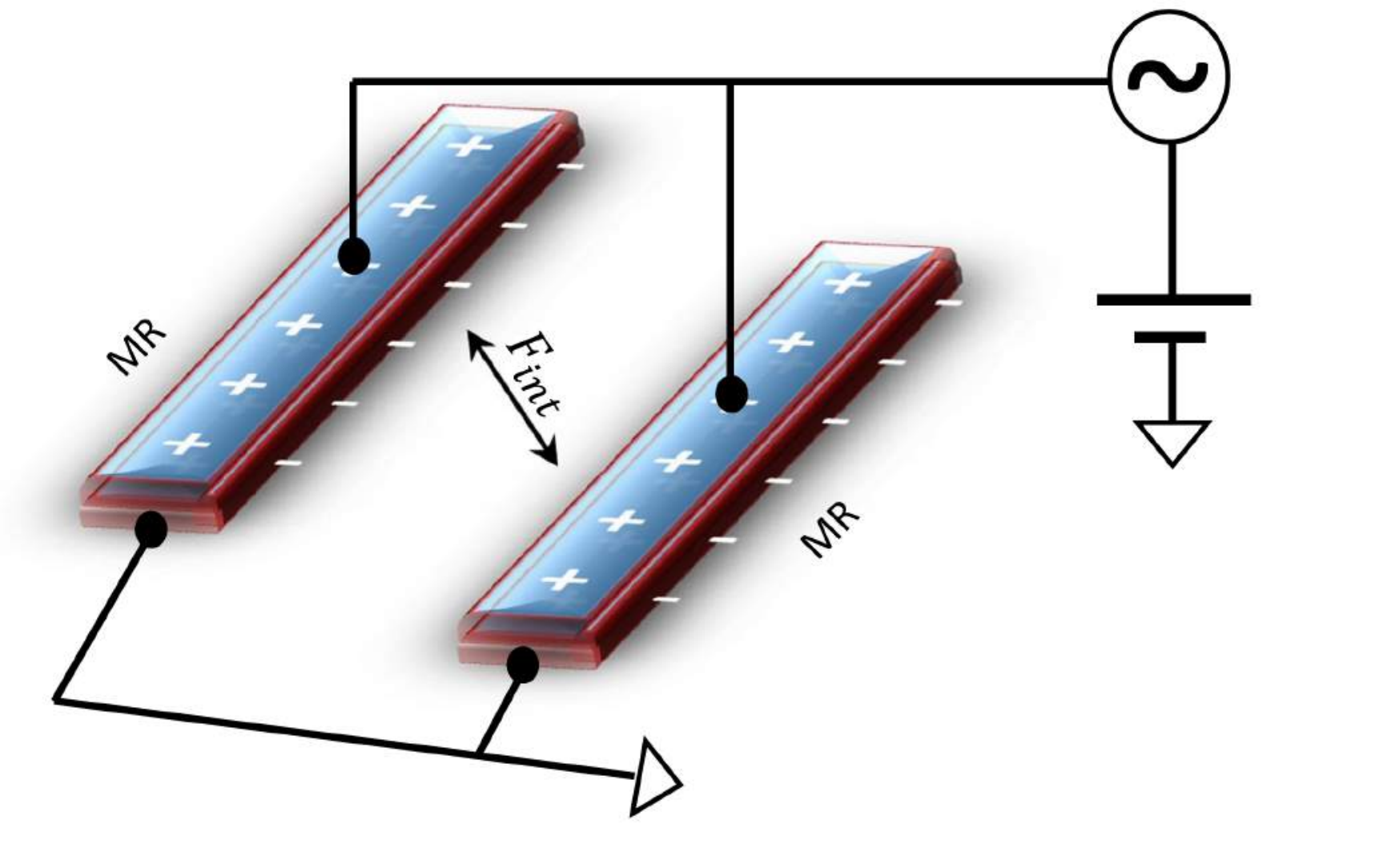}
\caption{~Schematic of an electrostatic coupling between two mechanical resonators. }
 \label{figbeamb}
\end{figure}
Different physical mechanisms may be responsible for an effective coupling between two nanomechanical resonators. The most important ones
are those of mechanical and electrostatic origin. A mechanical coupling occurs when resonators are
fabricated on the same chip and are connected to one another via an elastic mechanical structure. In this case, the
nature of the coupling and its strength strongly depends on the architecture of the nanomechanical
resonator system.

However, the coupling between two MRs can be realized also via the electrostatic coupling which is caused by the interaction of electric charges in the layers of the NMR devices. When the adjacent beams are polarized by an externally applied voltage, their top
and bottom layers are charged and as a result dipole moments are formed. In the case of a common top electrode~(see Fig. \ref{figbeamb}) the
dipole moments are identical, and thus experience a repulsive electrostatic force, which is given
by
\begin{equation}\label{force}
F_{int}= -G_0(X_1-X_2),
\end{equation}
where the coupling constant $ G_0 $ depends upon the distance between beams, the beams area, and the applied voltages. The electrostatic potential energy associated with the repulsive force~(\ref{force}) is given by
\begin{equation}\label{potential}
V_{int}= \frac{G_0}{2}(X_1-X_2)^2.
\end{equation}
Adding the electrostatic potential energy of Eq.~(\ref{potential}) to the nonlinear Hamiltonian of the beams Eq.~(\ref{Hamiltoniansingle}) gives the total Hamiltonian of the two coupled nonlinear resonators presented in the main text of the paper,
\begin{equation}\label{classicalham0}
H=\sum_{i=1,2}\Big[\frac{P^2_i}{2m_i}+\frac{m_i \omega_{0i}^2X^2_i}{2}\Big]+\sum_{i=1,2}\frac{\lambda_{0,i}}{4}X_i^4-G_0 X_1 X_2,
\end{equation}
where we have also redefined the effective frequency of the fundamental mode of each beam, $\omega_{01}, \omega_{02}$, by including also the (typically small) frequency shift associated with the dipole-dipole coupling constant $ G_0$.
\section{Number fluctuations in the Rabi and Josephson regimes}
In this section, we compare the phonon number fluctuations in the Rabi and Josephson regimes and we show that they are reduced in the Josephson regime. Note that complementary information can be found in Refs.~\cite{Gati,Paraoanu}.

We first introduce the relative number operator
\begin{equation}
\hat n=\frac{1}{2} (\hat b_1^{\dagger} \hat b_1-\hat b_2^{\dagger} \hat b_2 ),
\end{equation}
and recall that the total number of phonons $N_T\equiv\hat n_1+\hat n_2 $ is a constant of motion when damping is negligible. We can rewrite the annihilation operators $ b_1 $ and $ b_2 $ adopting the following polar decomposition
\begin{eqnarray}\label{anniop}
\hat b_1 &=& \sqrt{\frac{N_T}{2}+n}\,\,  \mathrm{e}^{-i \phi/2},\nonumber\\
\hat b_2 &=& \sqrt{\frac{N_T}{2}-n} \,\, \mathrm{e}^{i \phi/2}.
\end{eqnarray}
Substituting Eqs.~(\ref{anniop}) into the Hamiltonian of Eq.~(\ref{HamiltonianInt}), we get the phase representation for the Josephson Hamiltonian ($\hbar=1$)
\begin{equation}
H=\delta n+\frac{E_c}{2} n^2-E_j \sqrt{1-\frac{4n^2}{N_T} }\,\, \cos\phi,
\end{equation}
where $ \delta=[-\Delta_0+(N_T/2+1)(\lambda_1-\lambda_2)] $, $ E_c= \lambda_1+\lambda_2 $ is the charging energy, and $ E_j=G N_T $ is the Josephson coupling energy. In the limit of small phase oscillations and for $  \delta=0$, this effective Hamiltonian becomes
\begin{equation}\label{reducedHam}
H\simeq \frac{E'_c}{2} n^2+\frac{E_j}{2}\phi^2,
\end{equation}
\\
where $ E'_c=E_c+\frac{4E_j}{N^2_T} $ is the effective charging energy. Eq.~(\ref{reducedHam}) describes a harmonic oscillator with resonance frequency $ \omega=\sqrt{E'_c E_j} $ for which the root mean square of the number and phase fluctuations in its ground state can be expressed as
\begin{equation}\label{numberfluc}
\Delta n=\frac{1}{\sqrt{2}}\Big(\frac{E_j}{E'_c}\Big)^{1/4},
\end{equation}
\begin{equation}
\Delta \phi =\frac{1}{\sqrt{2}}\Big(\frac{E'_c}{E_j}\Big)^{1/4},
\end{equation}
and satisfy the minimum uncertainty relation $ \Delta n \Delta \phi=1/2 $.

Let us now focus onto the phonon number fluctuations in the Rabi and Josephson regimes. In the Rabi regime $E_c\ll \frac{4E_j}{N_T^2}  $ or equivalently $ g\equiv N_T(\lambda_1+\lambda_2)/4G\ll 1 $, and therefore the phonon number fluctuations in Eq.~(\ref{numberfluc}) can be written as
\begin{equation}
\Delta n_{R}\simeq \sqrt{N_T}/2.
\end{equation}
In the Josephson regime instead, we have  $ 1<g<N_T^2 $ so that $E'_c \simeq E_c = \lambda_1+\lambda_2 $, which gives the following phonon number fluctuation
\begin{equation}
\Delta n_{J}\simeq \frac{1}{\sqrt{2}}\Big(\frac{G\,N_T}{\lambda_1+\lambda_2}\Big)^{1/4}.
\end{equation}
Since in the Josephson regime $ 1\ll G N_T/\lambda_1+\lambda_2\ll N_T^2/4 $, we can conclude that the phonon number fluctuations in the Josephson regime are smaller than those in the Rabi regime i.e., $ \Delta n_{J}< \Delta n_{R}$.

\end{document}